\documentclass[reprint,aps,prl,superscriptaddress, floatfix]{revtex4-1}

\usepackage{graphicx}
\usepackage{dcolumn}
\usepackage{bm}
\usepackage{color}
\usepackage{xcolor}
\usepackage{epstopdf}
\usepackage{gensymb}
\usepackage{ulem}
\usepackage{float}
\usepackage{xr}
\usepackage{sidecap}
\usepackage{mathrsfs}
\usepackage{amsmath}
\usepackage{amsthm}
\usepackage{enumerate}
\usepackage{soul}
\usepackage{amssymb}
\usepackage{units}
\usepackage{todonotes}
\usepackage{verbatim}

\usepackage{verbatim}

\begin{document}

\preprint{APS/123-QED}

\title{Non-Local Phononic Crystals \\for Dispersion Customization and Undulation-point Dynamics\\}

\author{Arash Kazemi*}
\affiliation{Department of Mechanical Engineering, University of Utah, Salt Lake City, UT 84112, USA}

\author{Kshiteej J. Deshmukh*}
\affiliation{Department of Mathematics, University of Utah, Salt Lake City, UT 84112, USA}

\author{Fei Chen*}
\affiliation{Department of Mechanical Engineering, University of Utah, Salt Lake City, UT 84112, USA}

\author{Yunya Liu}
\affiliation{Department of Mechanical Engineering, University of Utah, Salt Lake City, UT 84112, USA}

\author{Bolei Deng}
\affiliation{Department of Electrical Engineering and Computer Science, Massachusetts Institute of Technology, Cambridge, MA 02139, USA}
\affiliation{Department of Mechanical Engineering, Massachusetts Institute of Technology, Cambridge, MA 02139, USA}

\author{Henry Chien Fu}
\affiliation{Department of Mechanical Engineering, University of Utah, Salt Lake City, UT 84112, USA}

\author{Pai Wang}
\affiliation{Department of Mechanical Engineering, University of Utah, Salt Lake City, UT 84112, USA}

\begin{abstract}
Dispersion relations govern wave behaviors, and tailoring them is a grand challenge in wave manipulation. We demonstrate inverse design of phononic dispersion using non-local interactions on one-dimensional spring-mass chains. For both single-band and double-band cases, we can achieve any valid dispersion curves with analytical precision. We further employ our method to design phononic crystals with multiple ordinary (roton/maxon) and higher-order (undulation) critical points and investigate their wave packet dynamics.
\end{abstract}
\maketitle

Phononic crystals and vibro-elastic metamaterials are architected heterogeneous solids for the manipulation of mechanical waves. They can exhibit many unconventional properties, such as frequency band gaps~\cite{liu2000locally,wang2015locally,konarski2020buckling,willey2022coiled,arretche2022physical,widstrand2022bandgap,ding2023thomson}, negative refraction~\cite{kaina2015negative,fang2021negative,danawe2022experimental,fang2022harnessing}, and topologically protected modes~\cite{wang2015topological,danawe2022finite,ding2022non,charara2022omnimodal}. They also have a wide range of applications in cloaking~\cite{cummer2007one,nassar2020polar,xu2020physical,martinez2022metamaterials,xu2022artificial}, signal manipulation~\cite{karki2021stopping,kruss2022nondispersive}, focusing~\cite{xu2022lamb,okudan2023improved} and energy trapping~\cite{de2020experimental,alshaqaq2022programmable}. Recently, by incorporating non-local (i.e., farther than nearest-neighbor) interactions~\cite{fleury2021non}, Rosa\,\&\,Ruzzene~\cite{rosa2020dynamics,nora2020APS,rosa2022small,rosa2022SPIE} demonstrated diffusive transport, and Wu\,\&\,Huang~\cite{wu2021micropolar,wu2022multifunctional} investigated active control, while 
Chen\,et\,al.~\cite{chen2021roton} showed roton-like dispersion~\cite{prada2008local,godfrin2012observation,iglesias2021experimental,schmidt2021roton,lyu2022detection,wang2022nonlocal}, where the local minimum of the dispersion curve resembles the roton behavior~\cite{landau1941theory,kolomeisky2022negative,chomaz2018observation} of the helium-4 superfluid~\cite{ feynman1954atomic,henshaw1961modes,bryan2018maxon,godfrin2021dispersion,muller2022critical} at low temperature. All these exotic and desirable dynamic behaviors hinge on the dispersion relation – how frequency depends on wave vector – that is intrinsic to each particular design. 
However, most studies so far have been focused on the forward problem from a given design to a set of band structures. It is a long-standing goal in the research community to solve the inverse problem from given dispersion bands to actual metamaterial designs so that exotic behaviors and functionalities can be realized on demand. Prior efforts to tailor specific dispersions~\cite{goh2019group,dong2022achromatic,jiang2022dispersion,wang2021efficient,arora2022deformation} or band gaps~\cite{isaacs2018inverse,goh2019inverse,goh2020inverse,morris2022expanding} typically relied on iterative searches with high computational costs, and they had only very limited success.
\begin{figure}[b!]
    \includegraphics[width=0.45\textwidth]{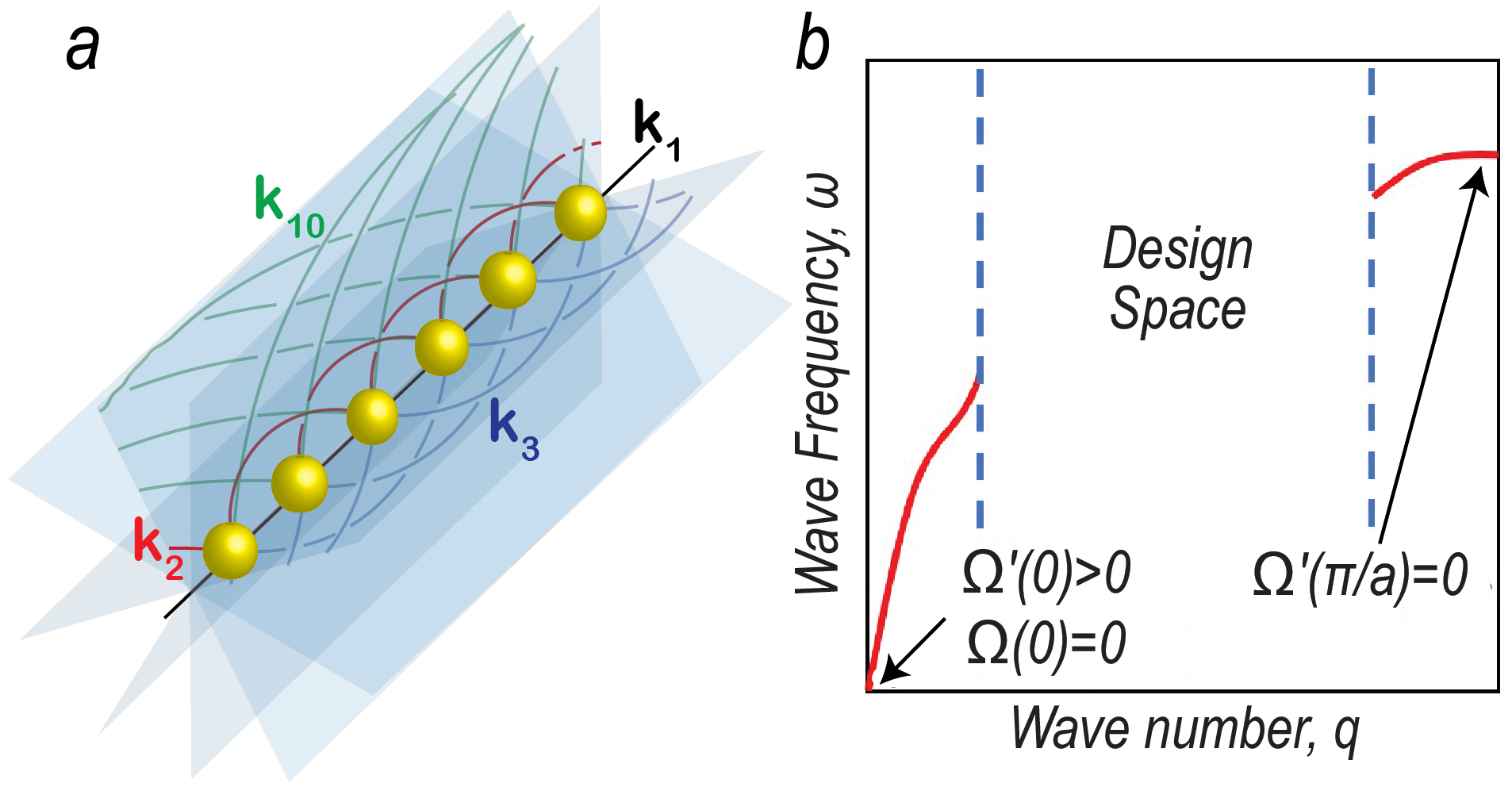}
    \caption{\label{fig:F1}
    (a) An infinite chain of identical masses. Each mass is connected to its $n$th-nearest neighbors with spring constant $k_n$. 
    (b) The design space with fundamental constraints at the center ($q=0$) and edge ($q=\pi/a$) of the $1^{st}$ Brillouin zone.}
\end{figure}
In this Letter, we demonstrate a design methodology that uses non-local interactions to customize dispersion relations. 
First, we present an analytical protocol to solve the inverse problem, achieving any arbitrarily-defined single-band dispersion on mono-atomic non-local chains. 
Then, we use this design protocol to obtain dispersion curves with ordinary and higher-order critical points. Using time-domain simulations, we illustrate their unconventional wave dynamics, especially at the undulation point (a.k.a. stationary inflection point), where both the first and second derivatives of the dispersion curve vanish. This results in highly concentrated vibration energy since the wave mode is simultaneously non-propagating and non-spreading.
Finally, we also investigate the di-atomic non-local chain and develop the design protocol to customize its two dispersion bands.

We start with a one-dimensional ``mono-atomic" phonon chain of identical masses, $m$, and linear springs.
A schematic of the model is depicted in Fig.\,\ref{fig:F1}(a). 
Each mass is connected with its two nearest neighbors by local interactions with the spring constant $k_1$. 
In addition, each mass is also connected on both sides to its two $n$th-nearest neighbors with non-local interactions specified by the spring constants $k_n$, for $n = 2, 3, 4, ...,N$, where $N$ is the longest-range non-local interaction in the system. 
The governing equation of motion for the $j^{th}$ mass is
\begin{equation}
\label{eq:motion}
m \ddot{u}_j = \sum_{n=1}^{N}  k_n(u_{j+n}-2u_j+u_{j-n}).
\end{equation}
Based on the Bloch theorem~\cite{hussein2014dynamics}, we obtain the following dispersion relation:
\begin{equation}
\label{eq:monodis}
\omega^2(q)  = \frac{2}{m} \Big(\sum_{n=1}^{N} k_n - \sum_{n=1}^{N} k_n \cos(nqa) \Big),
\end{equation}
where $\omega$ is the frequency, $q$ is the wavenumber, and $a$ is the spatial period of the lattice.
For conventional chains with local springs $k_1$ only, Eq.\,\eqref{eq:monodis} reduces to the classical result of $\omega^2(q)  = (4k_1/m)\sin^2{(qa/2)}$, which is always monotonic and reaches its maximum at the Brillouin zone boundary~\cite{phani2017dynamics}. 
The non-local interactions, on the other hand, may give rise to local minima and maxima at the interior of the Brillouin zone, as recently demonstrated by Chen\,et\,al.~\cite{chen2021roton} and earlier by Farzbod \& Leamy~\cite{farzbod2011analysis}. 
\begin{figure}[b!]
    \includegraphics[width=0.45\textwidth]{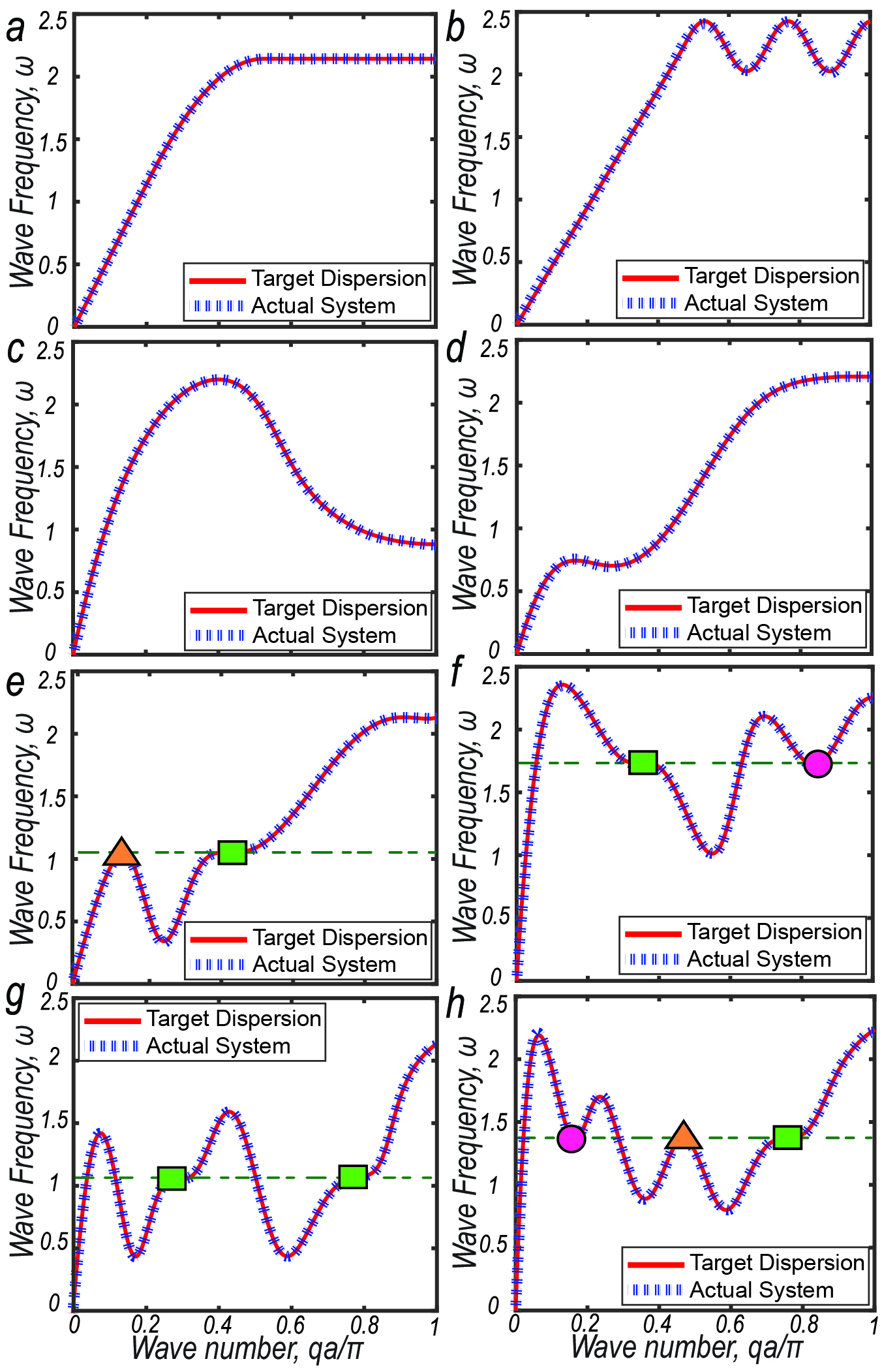}
    \caption{
    Customized dispersion curves with special features: (a) a flat top; (b)\,-\,(d) Non-monotonic behaviors at large, medium, and small wave number $q$ (i.e., at short, medium, and long wavelength as compared to lattice constant $a$), respectively; (e)\,-\,(f) maxons (triangles), rotons (circles), and undulation points (squares) occurring at the same frequency.}
    \label{fig:mono_examples}
\end{figure}

Since Eq.\,\eqref{eq:monodis} takes the form of a Fourier series, we can use it to tailor the non-local interactions to achieve any desirable dispersion behavior. Before the demonstration of customization procedures, it is necessary to understand all constraints in possible dispersion relations. Here, we consider the following physical and symmetry principles as fundamental assumptions of the designer non-local phononic crystals:\vspace{-0.1in}
\begin{itemize}
  \item Passive with no energy input or output.\vspace{-0.1in}
  \item Free-standing with no grounded springs.\vspace{-0.1in}
  \item Time-reversal symmetric with no gyroscopic effect.\vspace{-0.1in}
  \item Stable with a finite static stiffness.\vspace{-0.1in}
\end{itemize} 
Combining the above, we arrive at the requirements that, for any target dispersion relation $\Omega(q)$ defined on the non-negative half of the first Brillouin zone ($q \in [0, \pi/a]$) to be valid, it needs to be a smooth curve with (See Fig.\,\ref{fig:F1}(b)): 
\begin{equation}
\label{eq:symmcond}
\Omega(0) = 0, \quad
0< \Omega'(0)  < + \infty, 
\quad \text{and} \quad
\Omega'(\pi/a)  = 0.
\end{equation}
\noindent Given an arbitrarily specified dispersion relation, $\Omega(q)$, satisfying Eqs.\,\eqref{eq:symmcond}, we can design a non-local phononic crystal using the following protocol: First, we find the Fourier coefficients as
\begin{equation}
\label{eq:Fouriercoff2}
	A_{n}=\frac{2a}{\pi} \int_{0}^{\pi/a} \Omega^2(q) \cos(nqa)\,dq \text{,} \quad n = 1,2,...,N.
\end{equation}
Then, the design can be obtained by:
\begin{equation}
\label{eq:mono-design}
\begin{aligned}
k_{n}/m = - A_{n}/2 \text{,} \quad n = 1,2,...,N.
\end{aligned}
\end{equation}

Fig.\,\ref{fig:mono_examples} shows results of this protocol with several examples. Since Eq.\,\eqref{eq:mono-design} shows all $k_n$'s simply scale with $m$, we can set $m=1$ for all cases. In each case, we compare the target dispersion with the actual one by examining the normalized root mean square deviation (NRMSD) between them. We purposefully choose the target curves with various interesting features. In the implementation, we use analytical functions as the targets for Figs.\,\ref{fig:mono_examples}(a) and \ref{fig:mono_examples}(b). For other cases, we use piece-wise spline functions 
to construct target curves. The detailed procedures are given in the \textit{Supplemental Materials}~\cite{SI}.
For each target curve, the stiffness design variables are obtained using Eqs.\,\eqref{eq:Fouriercoff2} and \eqref{eq:mono-design}. The total number of stiffness types is $N$=10 for Figs.\,\ref{fig:mono_examples}(a)-\ref{fig:mono_examples}(d), $N$=20 for Figs.\,\ref{fig:mono_examples}(e)-\ref{fig:mono_examples}(g), and $N$=25 for Fig.\,\ref{fig:mono_examples}(h), respectively. 
The NRMSD is less than one percent in all cases, and the details are given in the \textit{Supplemental Materials}~\cite{SI}. We show that it is possible to achieve a flat band top [Fig.\,\ref{fig:mono_examples}(a)] as well as non-monotonic dispersion at relatively short [Fig.\,\ref{fig:mono_examples}(b)], medium - [Fig.\,\ref{fig:mono_examples}(c)], and long - [Fig.\,\ref{fig:mono_examples}(d)] wavelength regimes. In addition, for critical points on the dispersion, we can design systems where local maxima - maxons, local minima - rotons, and stationary inflection points - undulations can occur at the same frequency, as illustrated in Figs.\,\ref{fig:mono_examples}(e) and \ref{fig:mono_examples}(f).
\begin{figure}[b!]
\includegraphics[width=0.48\textwidth]{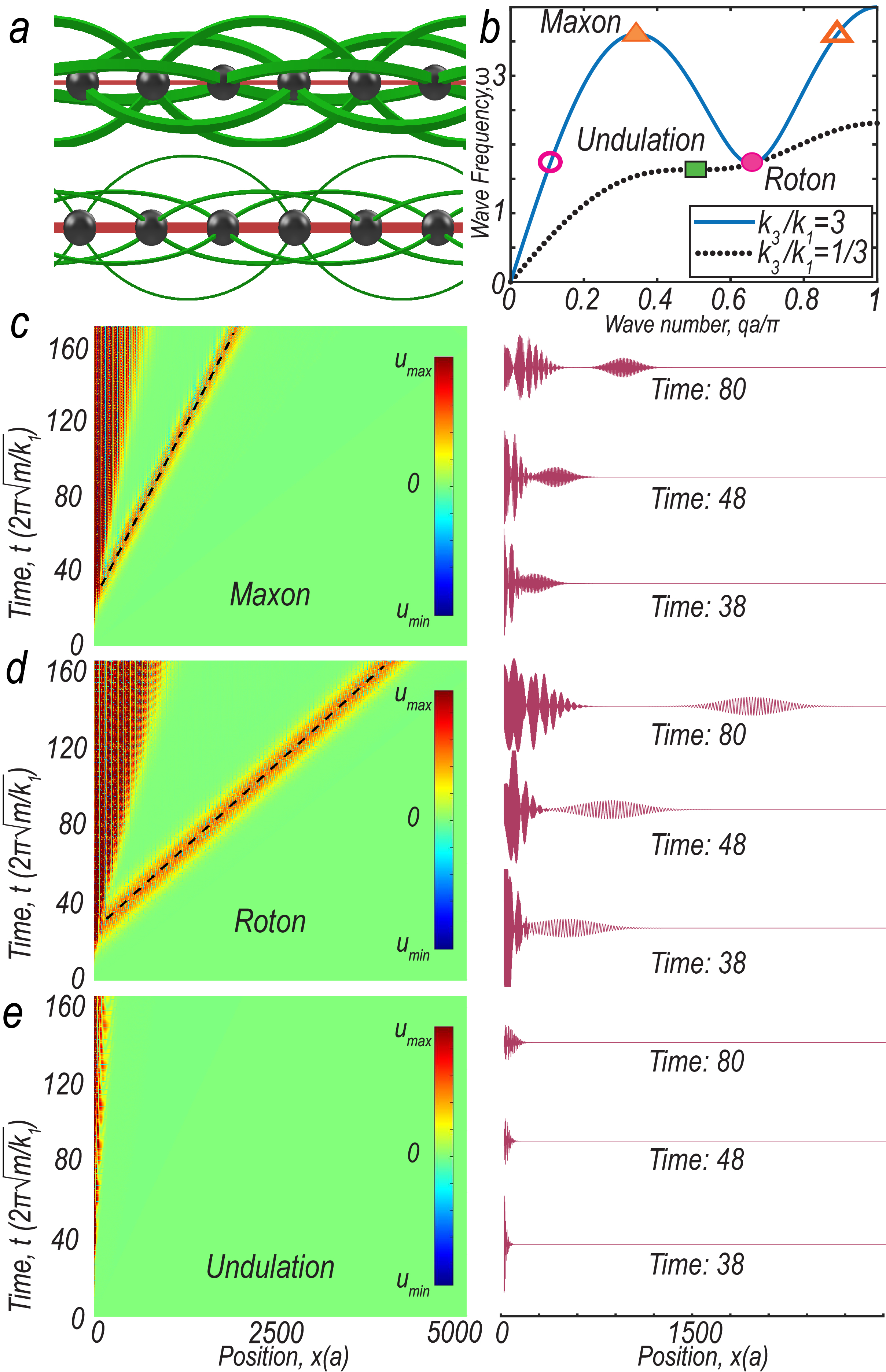}
\caption{\label{fig:time-domain}
(a) Schematics of two non-local phononic crystals with the first and third nearest neighbor interactions only - top:\,$k_3$=$3 k_1$, bottom:\,$k_3$=$k_1/3$.
(b) Dispersion curves: For $k_3$=$3 k_1$, a local maximum (maxon) appears at $(\omega, q a/\pi)$=$(3.61, 0.344)$, and a local minimum (roton) appears at $(\omega, q a/\pi)$=$(1.72,0.656)$.
For $k_3$=$k_1 / 3$, a stationary inflection point (undulation) appears at $(\omega, q a/\pi)$=$(1.63,0.5)$, where both the first and second derivatives vanish. (c)\,-\,(e) Time domain results for the 3 critical points in (b): maxon, roton, and undulation, respectively. The left column lists the time-space plots, while the right column shows wave amplitude snap-shots.}
\end{figure}

Next, we investigate these critical points that exhibit exotic dynamics by considering two specific instances of non-local phononic crystals with the third-nearest-neighbor ($k_3$) interactions as the only non-local effect. When, $k_3 = 3 k_1$ (Fig.\,\ref{fig:time-domain}(a)-top), the dispersion curve is non-monotonic (blue solid curve in Fig.\,\ref{fig:time-domain}(b)), exhibiting one local maximum (maxon-like) and one local miminum (roton-like~\cite{chen2021roton}) at $qa=2\tan^{-1}(\sqrt{\frac{11}{7}\pm \frac{6\sqrt{2}}{7}})$. Both of them represent critical-point wave modes with zero group velocity (ZGV), and they are analogous to the van Hove singularities~\cite{van1953occurrence} in electronic band structures. 
These ZGV modes also have promising applications in many wave-related engineering technologies such as non-invasive structural health monitoring~\cite{balogun2007simulation,laurent2015laser,geslain2016spatial,wu2022existence} since the highly localized wave modes can enhance both the vibration energy concentration and the signal-to-noise ratio in ultrasonic probing. 
In contrast, when $k_3 =  k_1/3$ (Fig.\,\ref{fig:time-domain}(a)-bottom), the dispersion curve is monotonic (black dotted curve in Fig.\,\ref{fig:time-domain}(b)) with an undulation point in the middle at $qa=\pi/2$, where both the first and second derivatives vanish. While roton-like dispersions were recently demonstrated~\cite{chen2021roton,prada2008local,godfrin2012observation,iglesias2021experimental,schmidt2021roton,lyu2022detection,wang2022nonlocal}, and undulation points of electromagnetic waves in optical waveguides were studied as frozen modes~\cite{ramezani2014unidirectional,li2017frozen,herrero2022frozen}, we show here, for the first time, a second-order-critical undulation point for vibro-elastic waves in phononic crystals.

To demonstrate wave behaviors at these critical points, we also perform two types of time-domain simulations on finite chains.

First, we apply a force excitation on the left most mass of a chain with $5000$ unit cells. The forcing function is a Gaussian envelope in time:
\begin{equation}
f(t) = \exp{[-{(t-t_\textrm{m})^2}/{\tau^2}]} \cos{(\omega_\textrm{c} t)},
\end{equation}
where $\omega_\textrm{c}$ is the carrier frequency corresponding to the critical point, $t_\textrm{m}$ is the peak time of the envelope, and $\tau = 100/\omega_\textrm{c}$ characterizes the time duration of the envelope.
Figs.\,\ref{fig:time-domain}(c) and \ref{fig:time-domain}(d) show the results for maxon-like and roton-like dynamics, respectively, in the chain with $k_3 = 3 k_1$.
In each case, two modes of the same frequency but different wavelengths are observed: One is the traveling mode (hollow triangle and circle in Fig.\,\ref{fig:time-domain}(b)) with finite group velocity, as indicated by the black dashed line, while the other is the ZGV mode (filled triangle and circle in Fig.\,\ref{fig:time-domain}(b)) localized at the source.
Although the maxon-like and roton-like ZGV modes are not traveling waves, the results show they do diffuse and spread out in space over time. 
In contrast, Fig.\,\ref{fig:time-domain}(e) shows the result at the undulation-point frequency on the chain with $k_3=k_1/3$.
Only one wave mode is observed. 
More importantly, not only is this mode non-propagating, but it is also non-spreading, as both the group velocity, $\omega'(q)$, and the diffusion rate, $\omega''(q)$, vanish. This is a unique feature that does not exist in ordinary ZGV modes.

\begin{figure}[t!]
\includegraphics[width=0.49\textwidth]{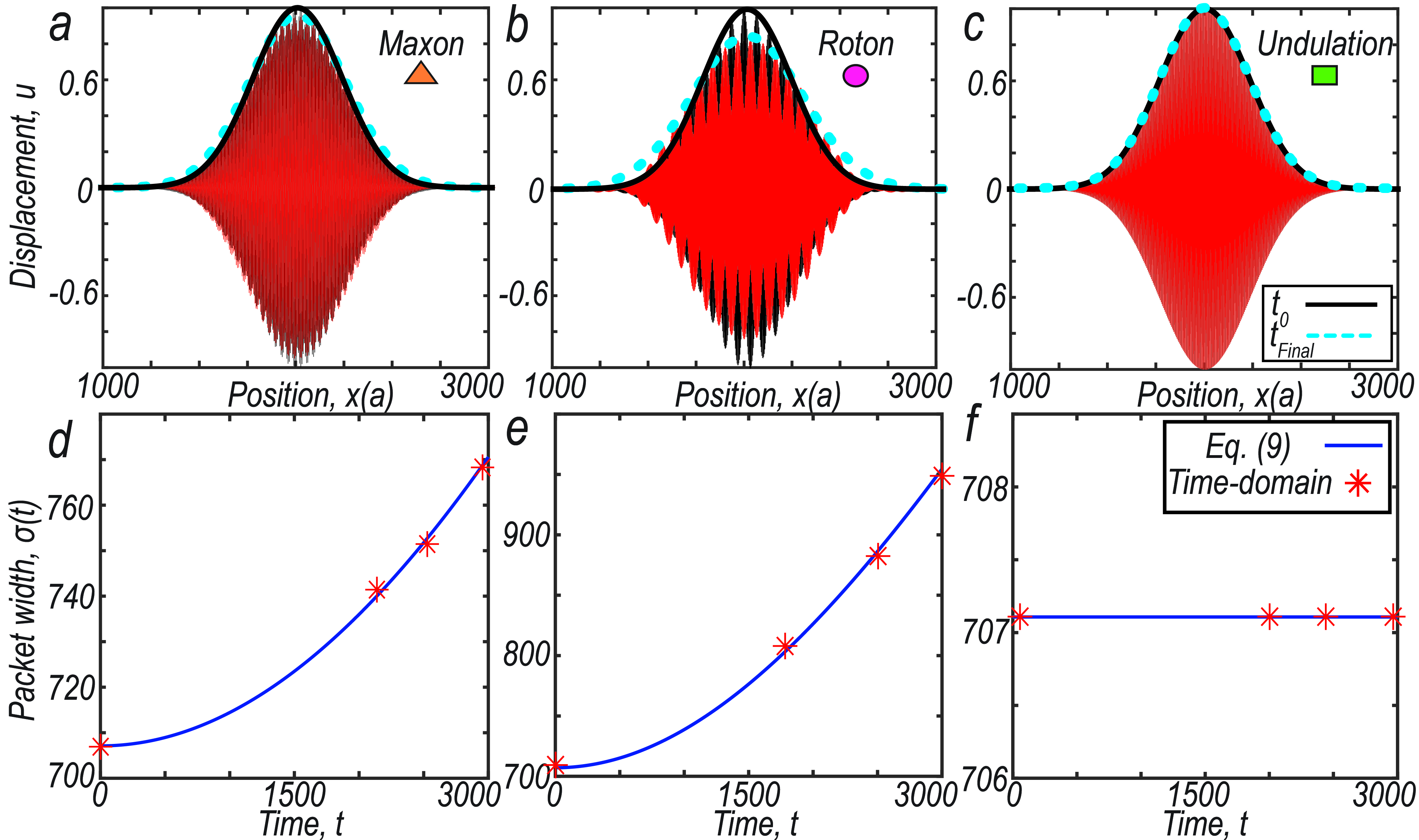}
\caption{
Diffusion at the critical points: Evolution of an initially prescribed Gaussian packet 
with different carrier wave number $q$ (i.e., different carrier wavelength) corresponding to the critical points in Fig.\,\ref{fig:time-domain}(b): (a) Maxon mode with $q a/\pi=0.656$; (b) Roton mode with $q a/\pi = 0.344$; (c) Undulation mode with $q a/\pi = 0.5$. 
The Gaussian envelope at time $t_{\textrm{Final}}$ (blue dotted curve) is compared to the initial envelope at $t_0$ (black solid curve).
(d)-(f) Theoretical and numerical values of $\sigma(t)$ vs. $t$, for each of the cases in (a)-(c) respectively, showing the spreading of wave envelopes.
} 
   \label{fig:Diffusion}
\end{figure}

\begin{figure}[t]
\includegraphics[width=0.5\textwidth]{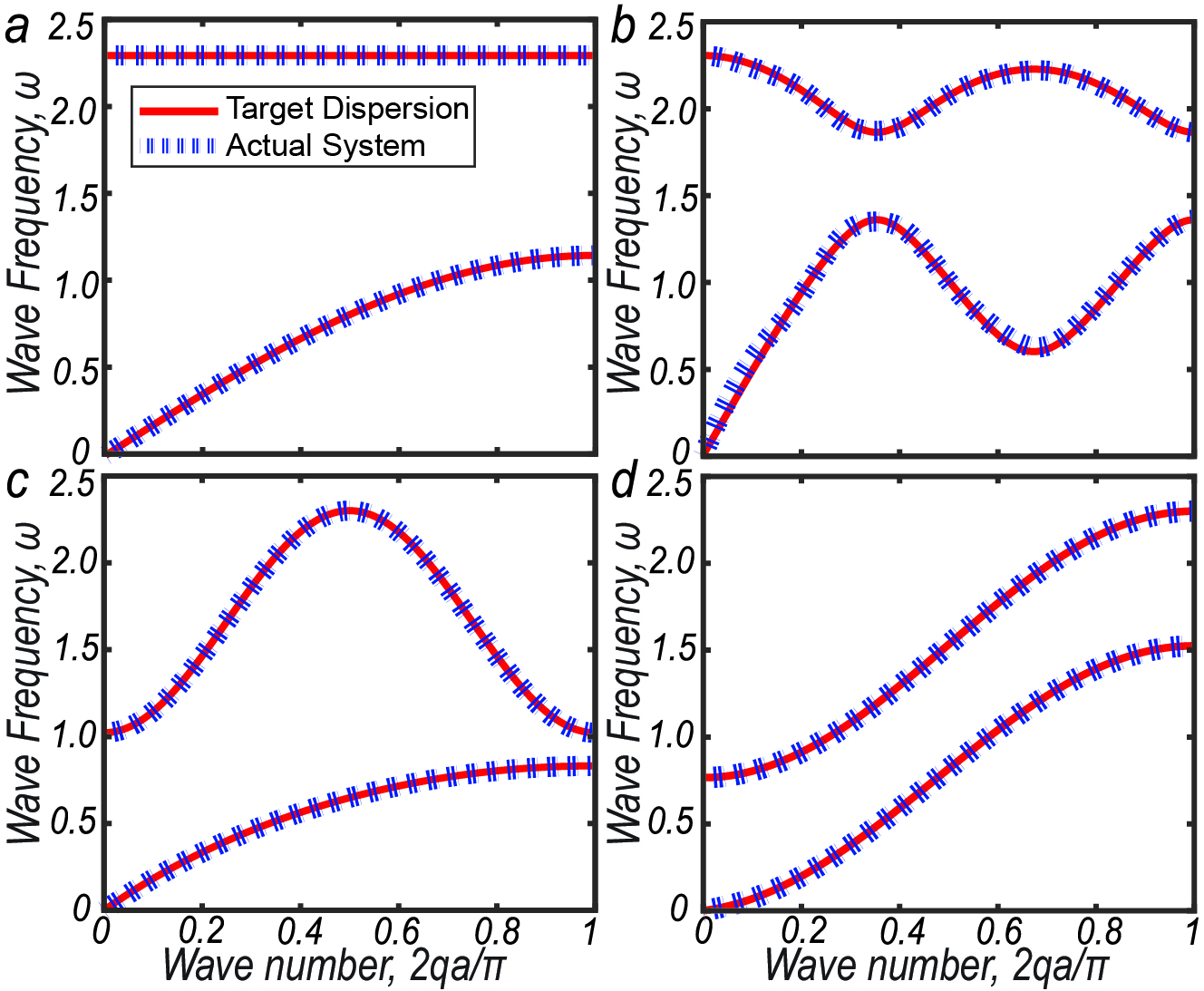}
\caption{\label{fig:F5}
Customized double-band dispersion relations.}
\end{figure}

Second, to further investigate the diffusion phenomena, we look into the time evolution of a localized Gaussian spatial wave packet,
\begin{equation}
u(x,t) = \exp{[-(x - x_0)^2/\sigma(t)]}\cos{q_\textrm{c} x},
\end{equation} 
where $q_c$ is the carrier wavenumber corresponding to the critical point, $x_0$ denotes the center of the wave packet, and $\sigma(t)$  characterizes the width of the envelope. We prescribe an initial Gaussian packet with $\sigma(t$=$0)$=$\sigma_0$.
In each case, there is only one wave mode associated with the prescribed wavenumber $q_c$ corresponding to the critical point, and it is a ZGV mode.
As such, the wave packet does not propagate. However, the wave packet can still spread out or diffuse in space, i.e. while maintaining the same mean $x_0$, the envelope width $\sigma(t)$ changes, and its evolution over time is governed by~\cite{remoissenet2013waves}
\begin{equation}\label{GVD}
    \sigma(t) 
    =
    \sigma_0 \sqrt{1 + \Big(t \omega^{\prime\prime} / {\sigma_0^2}\Big)^2}.
\end{equation}
Numerically, we can determine the diffusion rate of the wave packet by tracking $\sigma(t)$ in time-domain simulations on finite chains. 
Figs.\,\ref{fig:Diffusion}(a)-(c) show the comparison of wave packet diffusion for the $3$ critical points:  local maximum (maxon-like), local minimum (roton-like), and the undulation point (2nd order), respectively. 
In each case, the initial ($t=0$) wave envelope is represented by a  black solid line. After evolving for sufficient time ($t=t_\textrm{Final}$) the resulting wave envelope is shown as a blue dotted line. 
Figs.\,\ref{fig:Diffusion}(d)-(f) show, for each of the cases in Figs.\,\ref{fig:Diffusion}(a)-(c), respectively, the evolution of the packet width, $\sigma(t)$, at several time instances. 
Broadening of the envelope is observed for both maxon and roton packets, where $\omega^{\prime}=0$ but $\omega^{\prime\prime} \neq 0$. 
In contrast, the wave envelope preserves its initial shape without diffusion in the case of the undulation point, where $\omega^{\prime\prime}=\omega^{\prime}=0$.\\
Lastly, we also establish the customization protocol for the double-band system of a one-dimensional ``di-atomic" non-local phononic chain consisting of two different masses $m_1$ and $m_2$. This model leads to the following dispersion relations:
\begin{equation} \label{d2}
\begin{split}
\omega_{\pm}^2 &=  K_0(\frac{1}{m_1}+\frac{1}{m_2}) \\ &\mp \sqrt{K_0^2({\frac{1}{m_1} + \frac{1}{m_2}})^2 + \frac{1}{m_1m_2} (K_1^2 -4 K_0^2)},
\end{split}
\end{equation}
\noindent where $-$ and $+$ in the $\mp$ sign denote the first and second bands (historically referred to as ``acoustic" and ``optical" branches), respectively. Here, $K_{0}$ and $K_{1}$ are
\begin{equation}
\begin{split}
K_0 (q) &= \sum_{n=1}^{N} k_n - \sum_{{n=2 \atop{n \; \text{even}}}}^{N}  k_n\cos(n q a),\\
K_1 (q) &=  2 \sum_{{n=1 \atop{n \; \text{odd}}}}^{N} k_n  \cos{(nqa)},
\end{split}
\end{equation}
which are defined on the non-negative half of the first Brillouin zone, $q \in [0,\pi/(2a)]$. Given two arbitrarily specified smooth curves as the targets, $\Omega_-(q)$ and $\Omega_+(q)$, satisfying all fundamental and symmetry requirements detailed in the \textit{Supplemental Materials}~\cite{SI}, we can design a non-local chain using the following protocol: \\
First, we calculate 
\begin{equation}\label{get_alpha}
\begin{aligned}
\alpha &=  m_2/m_1 = {\Omega^2_{+}(\frac{\pi}{2a})} /{\Omega^2_{-}(\frac{\pi}{2a})},\\
A(q) &= [\Omega^2_{+}(q) + \Omega^2_{-}(q)] /{2}, \\
D(q) &= [\Omega^2_{+}(q) - \Omega^2_{-}(q)]/{2}.
\end{aligned}
\end{equation}
Then, we can get
\begin{equation}\label{get_K0_K1}
\begin{aligned}
K_{0}(q) &= \alpha A(q) /(\alpha + 1),\\
K^{2}_{1}(q) &= 4K^2_{0}(q)-\alpha A^2(q) + \alpha D^2(q),
\end{aligned}
\end{equation}
Lastly, we obtain the stiffness values as
\begin{equation}\label{eq_Diatomic_kn}
\begin{split} 
&k_n = \frac{2a}{\pi} \int_{0}^{ \frac{\pi}{2a}} K_{1} \cos(n a q)\,\mathrm{d}q
\text{,} \quad n = 1,3,5,... \\
&k_n =  - \frac{4a}{\pi} \int_{0}^{ \frac{\pi}{2a}} K_{0} \cos(n a q)\,\mathrm{d}q 
\text{,} \quad n = 2,4,6,...
\end{split}
\end{equation}
Fig.\,\ref{fig:F5} shows the results of this protocol with several examples by setting $m_{1} = 1$. The target curves are purposefully chosen with various features: Fig.\,\ref{fig:F5}(a) demonstrates a rising first band with a flat second band; Fig.\,\ref{fig:F5}(b) shows two bands with changing but always opposite convexity; Fig.\,\ref{fig:F5}(c) has a constant-curvature first band and an arched second band; and Fig.\,\ref{fig:F5}(d) has both bands monotonically increasing. These examples show that both localized and traveling wave modes can be designed at any arbitrarily desirable frequency and wavelength by our protocol on either band. In the implementation, we set the total number of stiffness types as $N$=$20$ for Fig.\,\ref{fig:F5}. We also examine NRMSD values between the target and actual dispersion curves.  The results show that, in most cases, a good match can be achieved with just a small number of non-local springs. Detailed information is summarized in \textit{Supplemental Materials}~\cite{SI}.

In conclusion, we can completely and analytically customize the dispersion relations in phononic crystals by incorporating non-local springs.
We show dispersion curves with multiple critical points of the first (maxon/roton) and second order (undulation). We further study the wave packet dynamics at each of the critical points and illustrate how we can use them to create novel behaviors of localized modes. This enables future research on higher-order critical points of elastic waves in terms of topology, scaling, and symmetry~\cite{yuan2020classification,chandrasekaran2020catastrophe} in 2D and 3D systems. Finally, we can also solve the inverse problem for arbitrary two-band dispersion relations.

For practical considerations, physical samples of phononic crystals with a small number of non-local springs can be realized in relatively simple designs~\cite{iglesias2021experimental,wang2022nonlocal}. We are confident that future research efforts may enable more sophisticated experimental setups with many more non-local interactions in higher dimensions.

At the continuum limit of the lattice constant $a\to 0$, wave mechanics in non-local continuum media can be described by higher-order strain-gradient models~\cite{askes2008four,deshmukh2022multiband,wang2022band} as well as \textit{peridynamics} \cite{silling2000reformulation, silling2010peridynamic}. In contrast to those popular phenomenological and semi-phenomenological approaches, our method has the advantage of prescribing system parameters to achieve desirable dynamic behaviors. Homogenizing our design methodology could potentially provide a route to design the micro-modulus elasticity kernel for target dispersion relations in continuum vibro-elastic metamaterials.

This work was supported by the P.W.'s start-up research funds of the Department of Mechanical Engineering at University of Utah.
K.J.D is grateful to the National Science Foundation for support through grant DMS-2107926. The support and resources from the Center for High Performance Computing at the University of Utah are gratefully acknowledged. The authors are grateful to Prof.\,Graeme Milton at Utah, Prof.\,Srikantha Phani at UBC, and Prof.\,Mahmoud Hussein at CU Boulder for inspirational discussions.

\bibliography{REF2}


\end{document}